\documentclass[conference]{IEEEtran}
\IEEEoverridecommandlockouts
\usepackage{cite}
\usepackage{amsmath,amssymb,amsfonts}
\usepackage{algorithmic}
\usepackage{graphicx}
\usepackage{textcomp}
\usepackage{xcolor}
\usepackage{booktabs}
\usepackage{tabularx}
\usepackage{multirow}
 \usepackage{hyperref}

\newcommand{\showComments}{yes}
\newcommand{\submit}{no}            

\newcommand{\note}[2]{
  \ifthenelse{\equal{\submit}{yes}}{}{%
    \ifthenelse{\equal{\showComments}{yes}}{\textcolor{#1}{#2}}{}
  }
}

\def\BibTeX{{\rm B\kern-.05em{\sc i\kern-.025em b}\kern-.08em
    T\kern-.1667em\lower.7ex\hbox{E}\kern-.125emX}}
\begin{document}

\title{\huge Automatic Data Transformation Using Large Language Model \\
-- An Experimental Study on Building Energy Data}

\author{\textcolor{black}{Ankita Sharma$^{a,b}$*, Xuanmao Li$^{b,g}$*, Hong Guan$^{a,b}$*, Guoxin Sun$^b$*, Liang Zhang$^{c,f}\ddag$,  Lanjun Wang$^d$,} 
\\ \textcolor{black}{Kesheng Wu$^a$, Lei Cao$^c$, Erkang Zhu$^e$, Alexander Sim$^a$, Teresa Wu$^b$, Jia Zou$^{a,b} \dag$}\vspace{3pt}\\
\textcolor{black}{Lawrence Berkeley National Lab$^a$, Arizona State University$^b$, University of Arizona$^c$, Tianjin University$^d$}\\ \textcolor{black}{Microsoft$^e$, National Renewable Energy Laboratory$^f$,  Huazhong Univ of Science and Technology$^g$ \thanks{* These authors made equal contributions; \dag  Jia Zou is the corresponding author; \ddag Liang Zhang is the contact for the datasets and use cases.}}
}


\maketitle

\begin{abstract}

Existing approaches to automatic data transformation are insufficient to meet the requirements in many real-world scenarios, such as the building sector. 
First, there is no convenient interface for domain experts to provide domain knowledge easily. Second, they require significant training data collection overheads. Third, the accuracy suffers from complicated schema changes.
To bridge this gap, we present a novel approach that leverages the unique capabilities of large language models (LLMs) in coding, complex reasoning, and zero-shot learning to generate SQL code that transforms the source datasets into the target datasets. We demonstrate the viability of this approach by designing an LLM-based framework, termed $\tt SQLMorpher$, which comprises a prompt generator that integrates the initial prompt with optional domain knowledge and historical patterns in external databases. It also implements an iterative prompt optimization mechanism that automatically improves the prompt based on flaw detection.
The key contributions of this work include (1) pioneering an end-to-end LLM-based solution for data transformation, (2) developing a benchmark dataset of $105$ real-world building energy data transformation problems, and (3) conducting an extensive empirical evaluation where our approach achieved $96\%$ accuracy in all $105$ problems.
${\tt SQLMorpher}$ demonstrates the effectiveness of utilizing LLMs in complex, domain-specific challenges, highlighting the potential of their potential to drive sustainable solutions.
\end{abstract}

\section{Introduction}
\label{sec:intro}

\begin{figure*} 
\centering 
\includegraphics[width=0.9\textwidth]{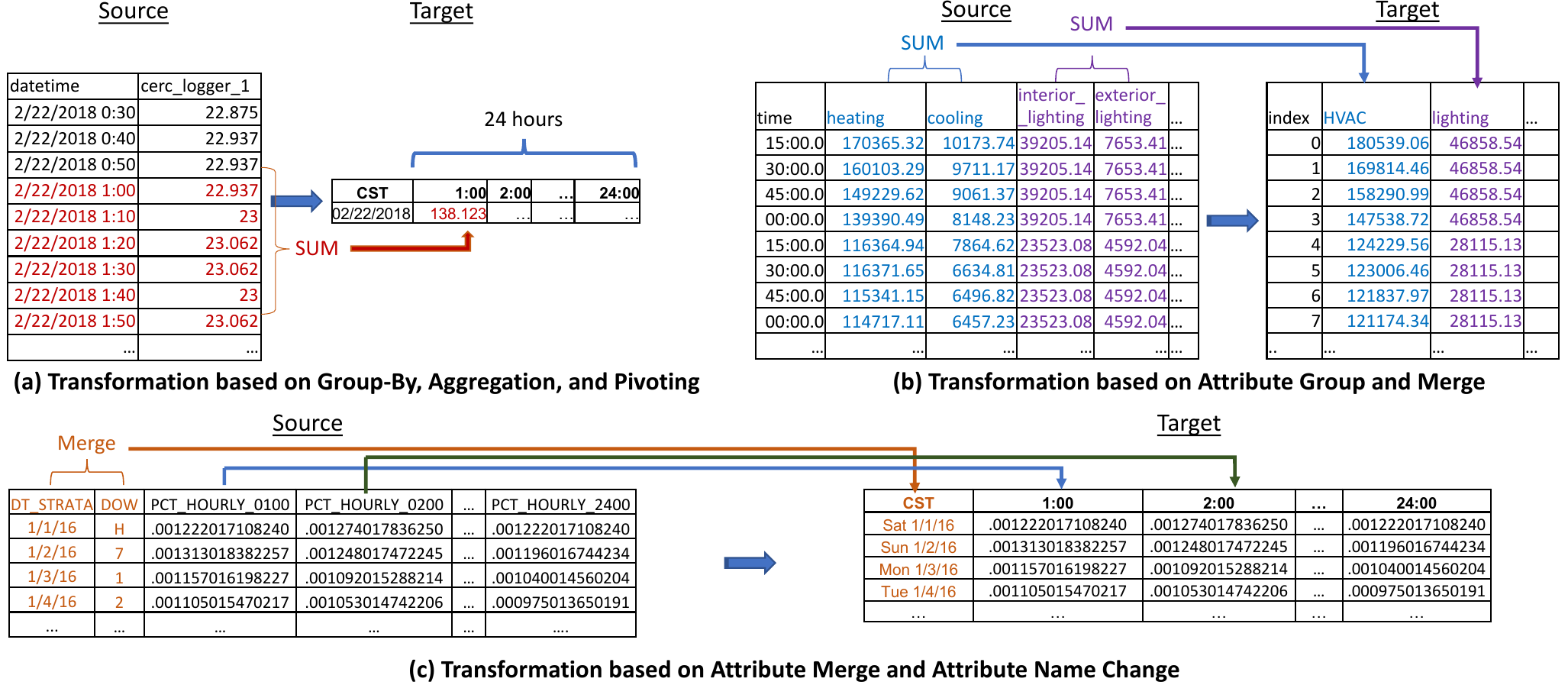}
\caption{\label{fig:overview} \small
Private sectors are using diverse formats to describe building load profiles. Each profile dataset must be converted into a unified target format for each different purpose. This figure provides several simplified examples.
}
\vspace{-10pt}
\end{figure*}

A recent study~\cite{energy2022} showed that in 2022, the end-use energy consumption by the building sector accounted for $\textbf{40}\%$ of total US energy consumption. It indicates that the energy management of buildings plays an important role in meeting the goals of energy sustainability~\cite{pinto2022transfer}. Automatic building energy management, including design, certification, compliance, real-time control, operation, and policy-making, requires the integration of data from diverse sources in both the private and public sectors. Harmonizing these data, as illustrated in Fig.~\ref{fig:overview}, remains a manual process. Extensive labor and expertise are thus required throughout the data lifecycle in the building sector. 
However, the state-of-the-art data transformation tools, such as Auto-Transform~\cite{jin2020auto}, Auto-Pipeline~\cite{yang2021auto}, and Auto-Tables~\cite{li2023auto}, are not effective due to the following gaps:

\vspace{3pt}
\noindent
$\bullet$ These tools are not publicly available and are based on supervised learning approaches, requiring non-trivial data labeling and training overheads.

 \noindent
 $\bullet$ The data transformation logic in the building sector involves multiple combinations of aggregation, attribute flattening, merging, pivoting, and renaming relationships between the source and the target. They are more complicated than existing data transformation benchmarks~\cite{li2023auto, yang2021auto}.  
 In addition, the accuracy achieved by the state-of-art tools on these simpler benchmarks is below $80$\% ~\cite{li2023auto, yang2021auto}, indicating it still requires human efforts to fix a significant portion of the cases. 
 
\vspace{1pt}
 \noindent
$\bullet$ Converting a building dataset to a target schema requires domain knowledge about both source/target schemas, which are available in domain-specific knowledge as illustrated in Fig.~\ref{fig:example1} and Fig.~\ref{fig:domain-specific-db-example}. However, there is no easy way to directly supply such knowledge in existing data transformation tools.


 \vspace{3pt}
To close the gaps, before this work, we once considered finetuning a pre-trained transformer model like BERT~\cite{devlin2018bert} to directly transform source data to target data~\cite{wang2020survive}. However, we identified many shortcomings of this approach. First, it is hard to formulate one unified predictive problem to transform data for all types of schema changes. Second, the transformation process is slow to handle large-scale data. Third, preparing a fine-tuning dataset for each task could also be challenging. 

This work proposes a novel and better approach, termed $\tt SQLMorpher$, which solves the problem in two steps. The first step is formulated as a Text2SQL problem~\cite{trummer2022codexdb, gu2023interleaving, gu2023few, popescu2022addressing}. We apply the LLM model to \textit{generate Structured Query Language (SQL) code that converts the source dataset(s) into the target dataset}. This step focuses on schema mapping, so we do not need to upload the entire source datasets. The second step applies the generated SQL code to efficiently transform the entire dataset in relational databases.

The idea is motivated by several key observations:
(1) \textbf{LLMs demonstrate superior performance in complex reasoning tasks}. In the building sector, domain experts often document the semantics of the source and the target tables in natural language. LLMs can better understand such descriptions and reason the relationships between the source and target than smaller pre-trained models.
(2) \textbf{LLMs have demonstrated strong coding and code explanation capability~\cite{chen2021evaluating}}. In addition, SQL's declarative nature makes it easier to map data transformation queries in natural language to SQL queries. 
(3) \textbf{LLMs have outstanding capabilities in zero-shot and few-shot adaptation and generalization.} 
Therefore, none or only a few training examples are needed.

\vspace{5pt}

Existing Text2SQL works~\cite{trummer2022codexdb, gu2023interleaving, gu2023few, popescu2022addressing} focus on selection queries, but cannot handle creation and modification queries. Furthermore, the utilization of LLMs for our target scenario is not only unique but is also faced with new challenges:

\noindent
\textbf{$\bullet$ Schema Change Challenge:} Different from existing Text2SQL works, $\tt SQLMorpher$ needs to generate the query that maps data from the source schema to the target schema.

\noindent
\textbf{$\bullet$ Prompt Engineering Challenge:} Designing a unified prompt to handle different types of schema changes and data transformation contexts is boring and tedious. 

\noindent
\textbf{$\bullet$ Accuracy Challenge:} Most importantly, the code generated by LLMs could be error-prone and even dangerous (e.g., leading to security concerns such as SQL injection attacks).

\begin{figure*} 
\centering
\includegraphics[width=1\textwidth]{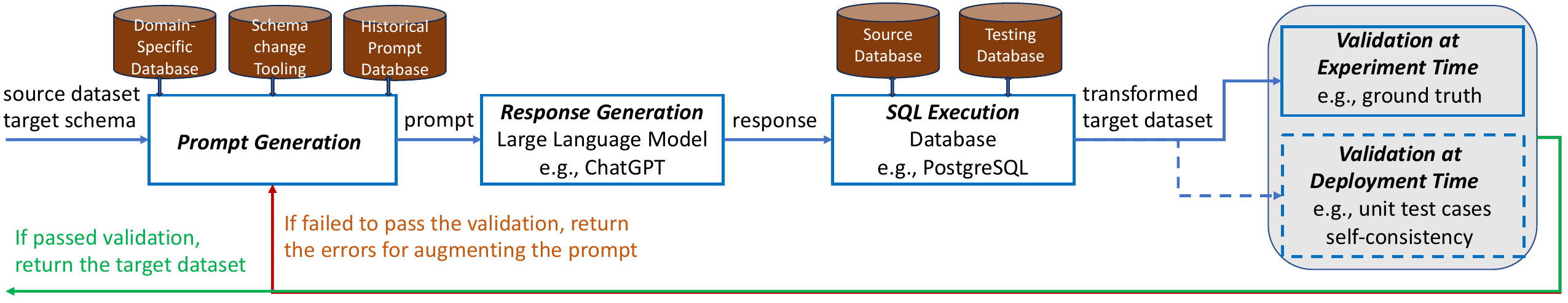}
\caption{\label{fig:workflow} \small
$\tt SQLMorpher$: Automatic Data Transformation based on LLM
}
\vspace{-10pt}
\end{figure*}

\vspace{3pt}
To address these challenges, the proposed system, as illustrated in Fig.~\ref{fig:workflow}, consists of the following unique components: 

First,  \textbf{a unique prompt generator} is designed to provide a unified prompt template. It allows external tools to be easily plugged into the component, such as domain-specific databases, vector databases that index historical successful prompts, and existing schema change detection tools ~\cite{fan2023semantics, dong2021efficient, wang2015schema} to retrieve various optional information. The prompt generator compresses the prompt size by using a few sample data to replace the source datasets for generating the SQL code applicable to transforming the entire source datasets.

Second,  \textbf{an automatic and iterative prompt optimization component} executes the SQL code extracted from the LLM response in a sandbox database that is separated from user data. It also automatically detects flaws in the last prompt and adds a request to fix the flaws in the new prompt. Examples of the flaws include errors mentioned in the last LLM response, the errors that occurred when executing the SQL query generated by the LLM, as well as insights extracted from these errors based on rules. 


\vspace{5pt}
\noindent
\textbf{Our Key Contributions} are summarized as follows: 


\noindent
\textbf{$\bullet$ } We are the first to apply LLMs to generate SQL code for data transformation. Our system, termed $\tt SQLMorpher$, includes a prompt generator that can be easily integrated with domain-specific knowledge, high-level schema-change hints, and historical prompt knowledge. It also includes an iterative prompt optimization tool that identifies flaws in the prompt for enhancement. We implemented an evaluation framework based on $\tt SQLMorpher$. (See details in Sec.~\ref{sec:framework})

\noindent
\textbf{$\bullet$ } We set up a benchmark that consists of $105$ real-world data transformation cases in $15$ groups in the smart building domain. We document each case using the source schema, the source data examples, the target schema, available domain-specific knowledge, the schema hints, and a working transformation SQL query for users to validate the solutions. We made the benchmark publicly available to benefit multiple communities in smart building, Text2SQL, and automatic data transformation~\footnote{\label{footnote:benchmark_repo}\url{https://github.com/asu-cactus/Data\_Transformation\_Benchmark}} \footnote{\label{footnote:code_repo}\url{https://github.com/asu-cactus/ChatGPTwithSQLscript}}. (See details in Sec.~\ref{sec:benchmark})

\noindent
\textbf{$\bullet$ } We have conducted a detailed empirical evaluation with ablation studies.  $\tt SQLMorpher$ using ChatGPT-3.5-turbo-16K achieved up to $\textbf{96}\%$ accuracy in $\textbf{105}$ real-world cases in the smart building domain. We verified that our approach can generalize to scenarios beyond building energy data, such as COVID-19 data and existing data transformation benchmarks. We also managed to compare $\tt SQLMorpher$ to state-of-the-art data transformation tools such as Auto-Pipeline (though these tools are not publicly available) on their commercial benchmark. The results showed that $\tt SQLMorpher$ can achieve $\textbf{81}\%$ without using any domain knowledge, and $\textbf{94}\%$ accuracy using domain knowledge, both of which outperform Auto-Pipeline's accuracy on this benchmark. We also summarized a list of insights and observations that are helpful to communities. (See details in Sec.~\ref{sec:eval})


\section{Related Works}

Existing Text2SQL tools~\cite{trummer2022codexdb, gu2023interleaving, gu2023few, popescu2022addressing} automatically generate SQL code to answer text-based questions on relations.  However, existing Text2SQL tools focus on generating selection queries. According to our knowledge, there do not exist any Text2SQL tools that support modification queries (e.g., insertions) that are required by data transformation. In addition, we surveyed multiple Text2SQL benchmarks including Spider~\cite{yu2018spider}, SQUALL~\cite{shi2020potential}, Criteria2SQL~\cite{yu2020dataset}, KaggleDBQA~\cite{lee2021kaggledbqa}, and so on. However, we didn't find any data transformation use cases in these benchmarks, which also indicates that data transformation problems are not the focus of today's Text2SQL research.

Existing automatic data transformation~\cite{abedjan2016dataxformer, gulwani2011automating, he2018transform, heer2015predictive, jin2018clx, singh2016blinkfill, jin2017foofah, zhu2017auto, jin2020auto, zuo2022spine, yang2021auto, li2023auto} fall in two categories: (1) Transform-by-Example (TBE)~\cite{abedjan2016dataxformer, gulwani2011automating, he2018transform, heer2015predictive, jin2018clx, singh2016blinkfill, jin2017foofah, zhu2017auto} infers transformation programs based on user-provided input/output examples, which have been incorporated into popular software such as Microsoft Excel, Power BI~\cite{becker2019microsoft}, and Trifacta~\cite{trifacta2020trifacta}. However, these works require users to provide examples of the transformed tuples, which is challenging for complicated data transformations.
(2) To address the issue, Transform-by-Target (TBT)~\cite{jin2020auto, yang2021auto, li2023auto} is recently proposed. Works in this category, such as Auto-Transform~\cite{jin2020auto}, Auto-Pipeline~\cite{yang2021auto}, and Auto-Tables~\cite{li2023auto}, transform data based only on input/output data schemas and optionally output data patterns. 
As mentioned, they learn a pipeline of data transformation operators using deep learning. They cannot easily integrate domain-specific knowledge represented in natural language or other formats. Although those tools are not publicly available, we conducted a comparison by running our approach on their benchmark, as detailed in Sec.~\ref{sec:non-smart-building}.

\section{$\tt SQLMorpher$ System Design}
\label{sec:framework}



As illustrated in Fig.~\ref{fig:workflow}, the $\tt SQLMorpher$ system consists of a prompt generator, a large language model (LLM), a SQL execution engine, and a component for iterative prompt optimization. In this section, we describe each component in detail. Although $\tt SQLMorpher$ is primarily engineered to evaluate the LLM in our target use scenarios, it is a first-of-a-kind design that has research values in defining the workflows and the interfaces between LLM and external tooling for the unique data transformation problem.

\subsection{Prompt Generation}
We designed a prompt template as illustrated in Fig.~\ref{fig:prompt}. The naive user must provide minimal information, such as the source and target table schemas and examples of the tuples in the source dataset.  Although a source table could contain many tuples, $\tt SQLMorpher$  only demonstrates to the LLM a few examples,  which are sufficient to generate code for correctly transforming the whole table. Despite the sampling techniques that can be applied here, we chose to randomly sample $5$ source tuples in the evaluation. If source tuples are not available, we asked the LLM to generate $5$ source tuples.

All other information is optional but is helpful for complicated transformation cases. We designed the prompt generator to retrieve additional information from external databases easily. Such information includes:

\begin{figure} [t]
\centering
\includegraphics[width=0.45\textwidth]{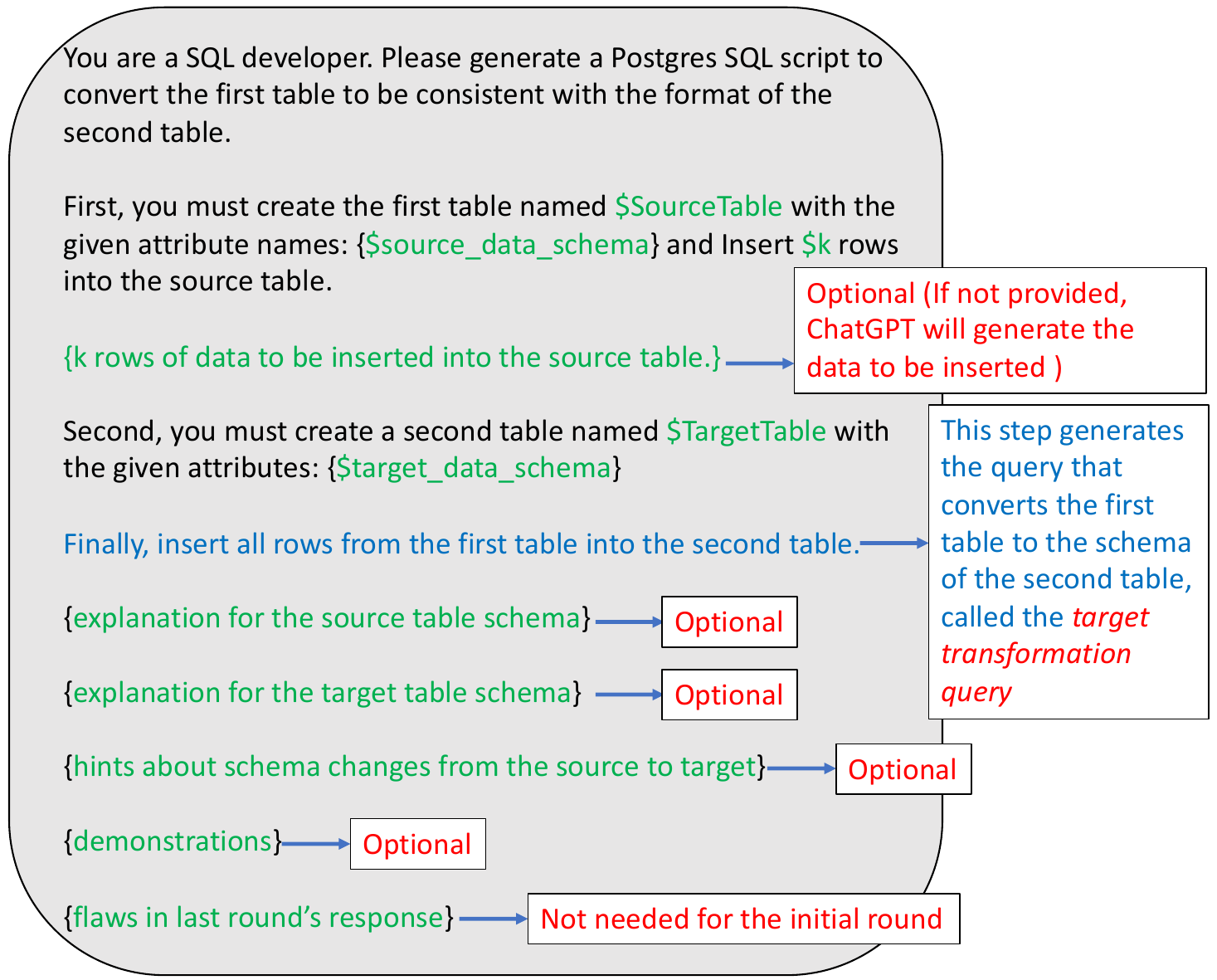}
\caption{\label{fig:prompt} \small
Prompt Template for single-table transformation.
}
\vspace{-10pt}
\end{figure}

\begin{figure*} [t]
\centering
\includegraphics[width=1\textwidth]{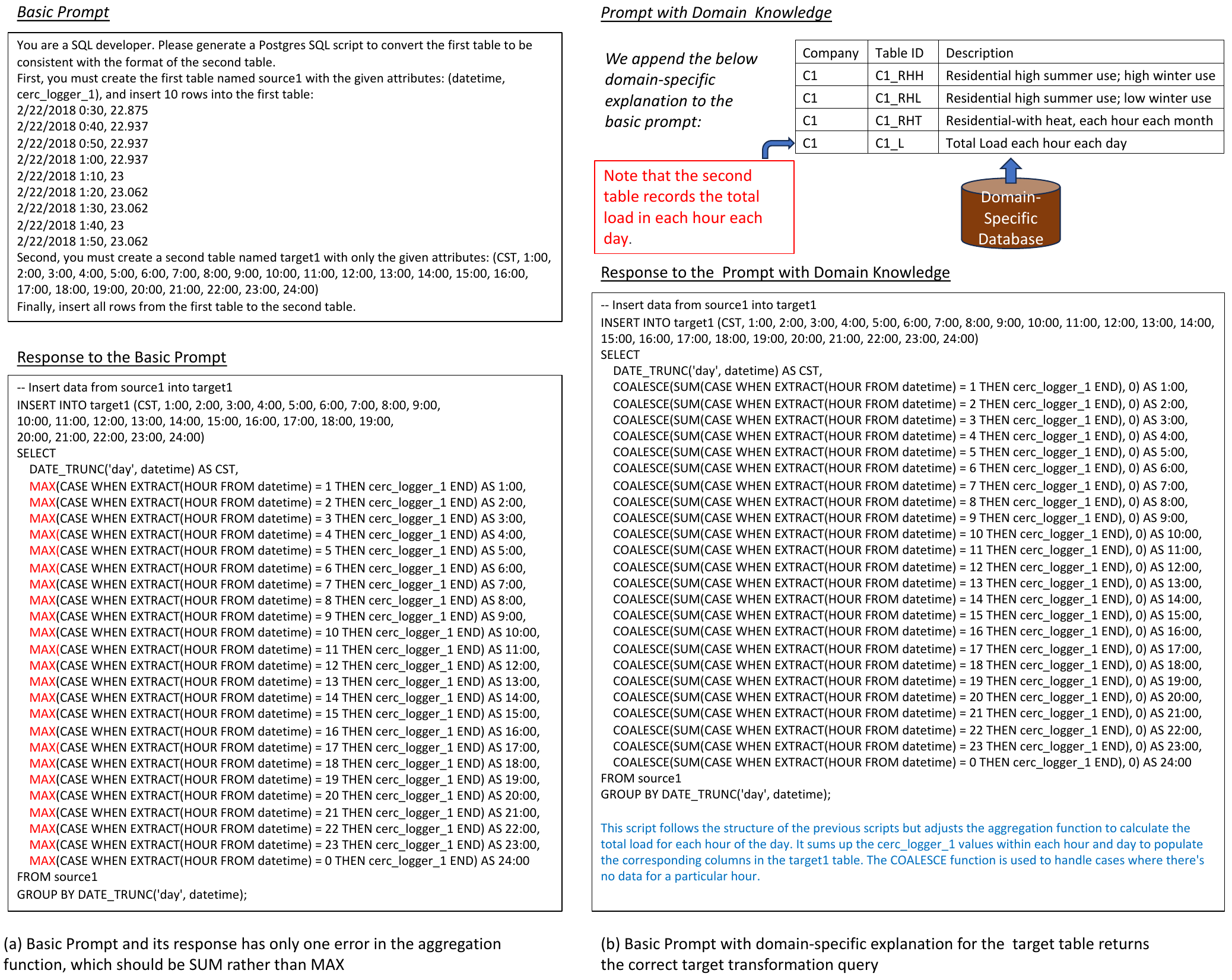}
\caption{\label{fig:example1} \small
Prompt-Response for the example illustrated in Fig.~\ref{fig:overview}a.
}
\vspace{-10pt}
\end{figure*}



\begin{figure*} [t]
\centering
\includegraphics[width=0.9\textwidth]{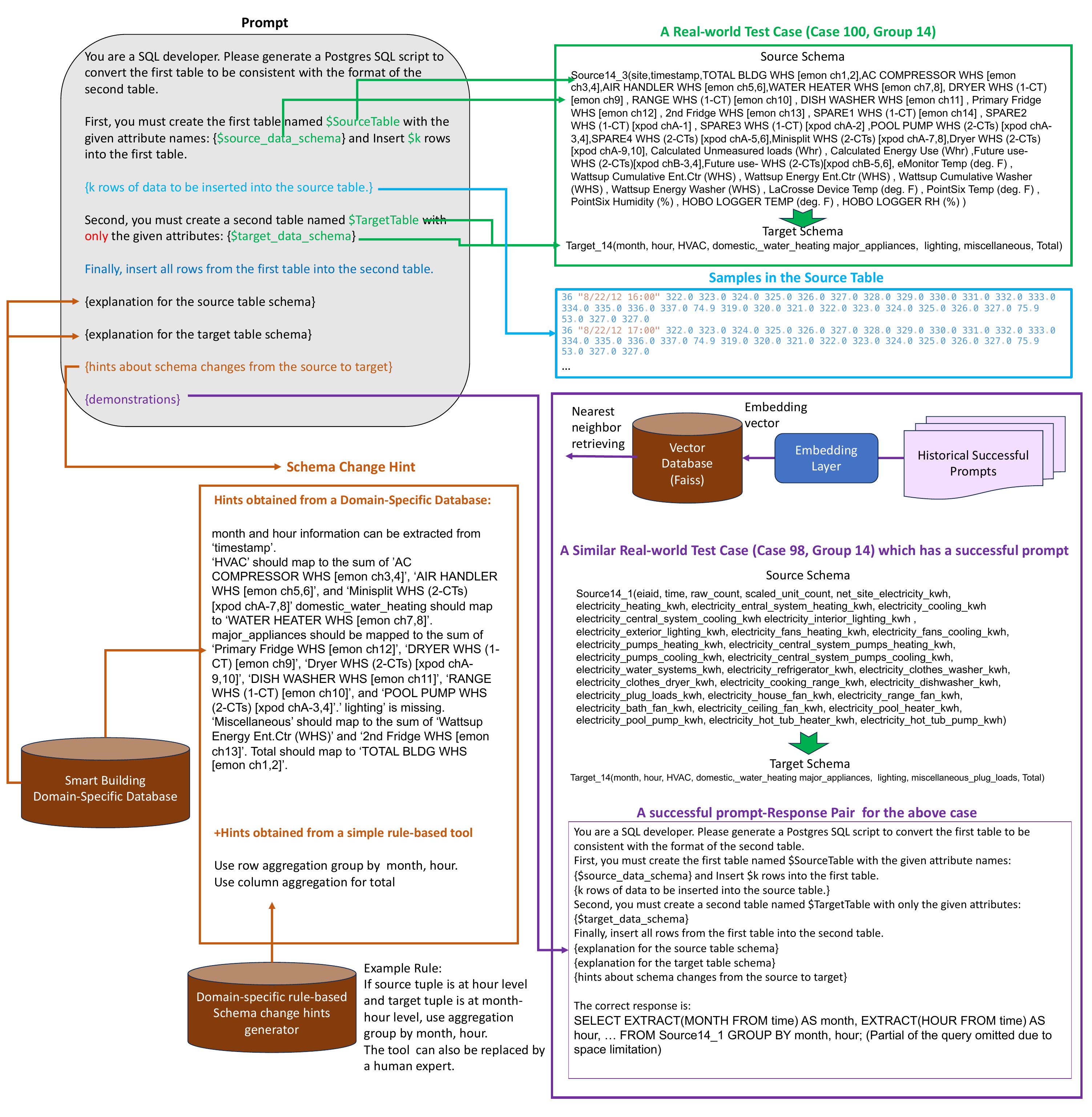}
\caption{\label{fig:example4} \small
A Working Prompt for a Real-World Case (Case 100 in Group 14 in Tab.~\ref{tab:benchmark}).
}
\vspace{-10pt}
\end{figure*}

\begin{figure} [t]
\centering
\includegraphics[width=0.45\textwidth]{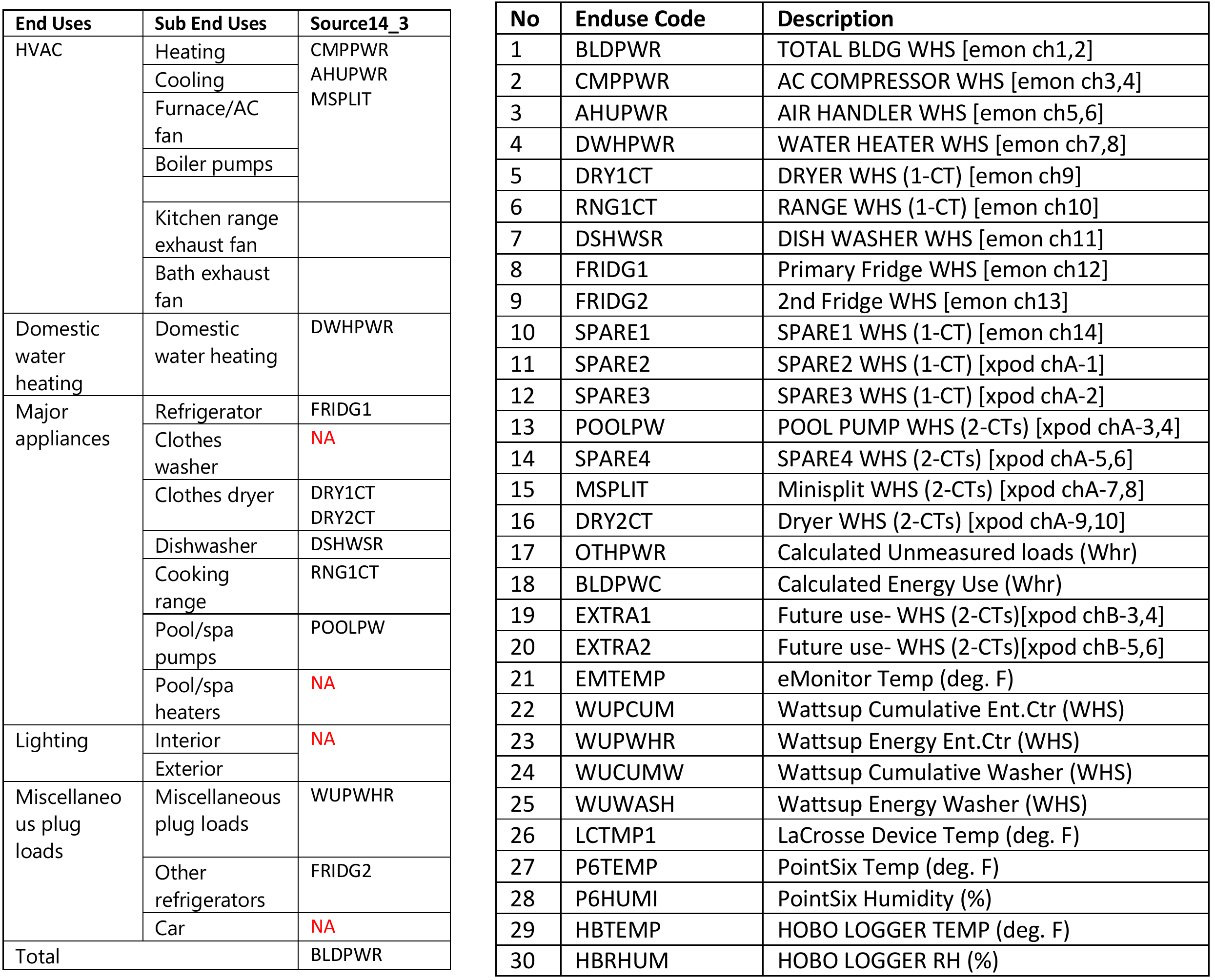}
\caption{\label{fig:domain-specific-db-example} \small
Example information from the smart building domain-specific database specifies the mapping from source attributes to the target attributes for the example in Fig.~\ref{fig:example4}.
}
\vspace{-10pt}
\end{figure}

\noindent
(1) \textbf{Domain-specific information}, which explains the semantics of each attribute in the source table and the target table. Given LLMs' diverse and ocean-volume training corpus, such explanations are not required for many domains, and so it is marked as optional. However, we found that using domain knowledge to enhance the prompt could be critical for many smart building data transformation cases. This information can be retrieved from a domain-specific database, as illustrated in Fig.~\ref{fig:example1}. In this example, basic information plus the domain-specific information that explains only the target table is sufficient to address the first example, as shown in Fig.~\ref{fig:overview}a.

\noindent
(2) \textbf{Schema change hints} suggests how the source schema is mapped to the target schema. Given the strong semantic reasoning capability of LLMs, hints are also optional. We found that some high-level hints, such as ``use aggregation", are sufficient for LLMs to generate correct Group-By clause and aggregation functions in most scenarios. As illustrated in the brown box in Fig.~\ref{fig:example4}, such information can be provided by (a) a rule engine that analyzes domain-specific databases as illustrated in Fig.~\ref{fig:domain-specific-db-example}, (b) a schema mapping tool such as Starmie~\cite{fan2023semantics}, or (c) even an LLM itself (e.g., using a separate LLM prompt that asks the LLM to identify schema changes between the source and the target). Fig.~\ref{fig:domain-specific-db-example} illustrates the example information that is available in a domain-specific database for smart buildings that can be leveraged to generate schema change hints. In this experimental study, most schema change hints are derived from the domain-specific databases as illustrated in Fig.~\ref{fig:example1} and Fig.~\ref{fig:domain-specific-db-example}.

\noindent
(3) \textbf{Demonstrations} add a few examples of historical prompt-response pairs to the prompt to perform few-shot learning. It is critical that the demonstrating prompts need to be similar to the current prompt. For $\tt SQLMorpher$ design, we choose to store the embedding vectors of historically successful prompts in a vector database, such as Faiss for top-$k$ nearest-neighbor search, as illustrated in Fig.~\ref{fig:example4}. %
In this example, the prompt will fail unless it includes both the schema change hints (in the brown box) and the demonstration (in the purple box).
We used the ChatGPT 3.5-turbo-16k model API in August 2023 to generate all examples in this section.

To retrieve the various types of information to augment the prompt, the $\tt SQLMorpher$ design includes a callback system. Each type of information corresponds to an event, and the user can register one or more callback functions with an event. Each callback function is expected to return a JSON object that specifies the retrieved information as well as a status code and error message that specifies connection or execution errors, if any. When generating a prompt, $\tt SQLMorpher$ will go through all types of information, and invoke all callback functions associated with each information type.


\subsection{SQL Execution}
\label{sec:sql_exec}


Compared to existing Text2SQL that focuses on selection queries that are read-only, leveraging LLM to generate modification queries is more complicated, partially because running the generated query may raise security concerns. 
In the initial iteration for a given user request, the system automatically duplicates the source dataset in a separate PostgreSQL database that serves as a sandbox environment to isolate the errors, if the duplicate does not exist. This is to ensure that the generated code will not corrupt the source dataset.  Then, the script creates the target table. Finally, it runs the generated query to transform the entire source dataset into the target format and insert all transformed tuples into the target table. If another iteration is needed, e.g., the response cannot pass the validation tests, the target table will be removed or archived before running the next iteration. 

\subsection{Validation}
\label{sec:validation}
The validation in the production environment could be challenging due to the lack of ground truth. It needs an automatic quality measurement (e.g., unit test cases, self-consistency, or accuracy of downstream tasks) for the transformed data, which we leave for future work to address.

In this work, we manually prepare the ground truth transformation queries for each transformation case in the experimental environment. At the validation stage, the ground truth transformation query will be executed against the source table, resulting in the target table, which is called the \textit{ground truth target table}. At the same time, by executing the target transformation query contained in the LLM response, as described in Sec.~\ref{sec:sql_exec}, we can also obtain a target table, which is called the \textit{generated target table}. 

We designed a validation script, which compares the \textit{generated target table} to the \textit{ground truth target table}. The comparison first validates whether two tables have the same number of attributes and tuples. Then, it performs attribute-reordering and tuple-sorting to ensure two tables share the same column-wise and row-wise orderings. Furthermore, the script will compare the similarity of the values for each attribute in the two tables. We use the ratio of the number of equivalent values (difference should be less than $e^{-10}$) to the total number of values to measure the similarity of numerical attributes. We use the Jaccard similarity to measure the similarity of categorical and text attributes. We average the similarity for each attribute to derive an overall similarity score. If the similarity score is below $1$, the validation fails.


\subsection{Iterative Prompt Optimization}
\label{sec:iterative}


This component is incorporated to evaluate the LLM's potential self-optimization capability for the data transformation problem.
If the validation fails, the prompt will be automatically augmented by identifying errors in the prompt: (1) errors mentioned in the LLM response or met when executing the generated transformation query; (2) errors detected in the transformed dataset, e.g., reporting the difference between the schema of the transformed dataset and the target schema; (3) inconsistency between the schema change hint and the response query, e.g., reporting if the hint specifies to use aggregation, but no Group-By or aggregation functions have been used.
Then, these errors will be appended to the prompt, and the new prompt will be sent back to the LLM, and it will repeat this process until it passes the validation, the maximum number of iterations has been reached, or the new prompt has no difference with the last prompt.

An example of \textit{useful} prompt flaws that we observed in those cases is ``ERROR:  INSERT has more expressions than target columns LINE 100: PCT\_HOURLY\_2500". Before adding this error to the prompt, ChatGPT cannot correctly handle an attribute that exists in the source table but not in the target table, PCT\_HOURLY\_2500. Adding the error to the prompt will resolve the problem.


\section{Experimental Evaluation}
\label{sec:eval}
In this section, we first describe the goal of the comparison study and all baselines that were used. Then, we present a benchmark, which is the \textit{first} benchmark for smart building data standardization problems. We further describe the setup of the experiments and the evaluation metrics. Ultimately, we will present and analyze the results and summarize key findings.

\subsection{Comparison and Baselines}

In this work, we mainly compare the effectiveness of six different types of initial prompt templates: 

\begin{itemize}
\item Prompt-1: Basic prompt with domain-specific description for the target schema.

\item Prompt-2: Prompt-1 with a domain-specific description for the source schema.

\item Prompt-3: Prompt-2 with schema change hints.

\item Prompt-1+Demo: Prompt-1 with one demonstration.

\item Prompt-2+Demo: Prompt-2 with one demonstration.

\item Prompt-3+Demo: Prompt-3 with one demonstration.
\end{itemize}

The first three prompt templates are designed for zero-shot learning when there does not exist a database of abundant historical working prompts. The last three prompt templates are designed for one-shot learning.

We also considered comparing our approach to Auto-Pipeline~\cite{yang2021auto}, which is a state-of-the-art automatic data transformation tool that only requires schema and tuple examples of the source and target tables and applies deep reinforcement learning to synthesize the transformation pipeline. 

\subsection{Benchmark Design}
\label{sec:benchmark}

\subsubsection{Building Energy Data Transformation}
\label{building-energy-data-transformation}
We collected $105$ data transformation examples in the smart building domain from $21$ energy companies in the United States. These examples are divided into $15$ groups so that each group has one target dataset and multiple source datasets of different types. Each source needs to be converted to the target format in the group. The groups are described in Tab.~\ref{tab:benchmark}. In Tab.~\ref{tab:smart-building-zero-shot-results}, we further show more statistics of the $105$ test cases by groups: (1) the number of distinct SQL keywords used in the ground truth query and (2) the length (i.e., number of characters) of the ground truth query. For each group, we compute the average of the above metrics for all cases in the group.

We document the following information in the benchmark: (1) target schema and domain-specific explanations for attributes; (2) for each source dataset, its schema, domain-specific explanation of attributes, examples of instances, schema change hints for transforming the source table to the target format, and the ground truth query that transforms the source to the target. The benchmark dataset is open-sourced in a GitHub repository \footref{footnote:benchmark_repo}.  

\begin{table*}[t]
    \centering
    \scriptsize
    \caption{\label{tab:benchmark} Descriptions of Benchmark Groups}
    \begin{tabular}{rp{2.5cm}p{3.5cm}p{4.5cm}cp{4cm}}
        \toprule
       ID & Group Description & Target Schema & Target Description & \#Sources & Source Descriptions \\
        \midrule
        1 &\multirow{6}{1.5cm}{Daily Hour-Level Load Profile Transformation} & Date 1:00 2:00 ... 24:00 & Date is of the format DOW MM/DD/YY, such as 'Fri 01/01/2016' & 10 & \multirow{8}{4cm}{Sources include load profiles captured per minute, per 5-minute, per 10-minute and per hour with different schemas and column names. Some example source schemas are as follows: Ex1. (DateTime LoadValue), where DateTime is a timestamp such as '2/22/2018 0:30'. Ex2. (Segment Date 1:00 2:00 ... 24:00:00), where Date is in the format of DOW MM/DD/YY such as 'Wed 01/01/2003', and Segment is an attribute that should be discarded from the relation. }\\ \cline{3-5}

        2 &  &DT DOW HOURLY\_0100 HOURLY\_0200 ... HOURLY\_2400 HOURLY\_2500& DT is in the MM/DD/YY format. DOW has values from 1-7 corresponding to Mon-Sun. Hourly\_2500 is used for leap second. & 10  &  \\ \cline{3-5}

        3 &  &Date Hour1 Hour2 ... Hour24& Date is of the format of MM/DD/YYYY & 10 &  \\ \cline{3-5}

        4 &  &Date 1:00AM 2:00AM ... 12:00PM& Date is of the format of MM/DD/YYYY & 10  &  \\ \cline{3-5}

        5 &  &Date Value1 Value2 ... Value24& Date is of the format of MM/DD/YYYY & 10 &   \\ \cline{3-5}

        6 &  &Date Hr1 Hr2 ... Hr24& Date is of the format of MM/DD/YYYY & 10 &  \\ 

          \cline{1-5}

        7 & \multirow{2}{1.5cm}{Monthly Hour-Level Load Profile Transformation} & Month DayType HR1 HR2 ... HR24 & Month has values such as January, Feburary, etc. DayType can be weekday or weekend. & 10 &  \\ \cline{3-5}

        8 &  & Hour January February ... December & Hour has values from 1 to 24. January records the average load in the corresponding hour in January. Other columns are similar. & 10 &  \\ 

          \midrule

        9 & Seasonal Temperature Range Transformation &  Season DayType Hour Temperature\_Range Constant Coefficient Low\_End High\_End & Season has values such as SPRING, SUMMER, FALL, and WINTER. Datatype can be either WEEKDAY or WEEKEND. Low\_End is the lowest temperature. High\_End is the highest temperature.  & 3 &  Source 1 and 2 are hourly temperature data grouped in four ranges and five ranges, respectively. Source 3 is seasonal temperature data grouped in three ranges.\\

         \midrule

        10 & Daily Hour-Level Load Transformation by Detailed Enduse & Datetime HVAC water\_heating Refrigerator Clothes\_washer Clothes\_dryer Dishwasher Cooking\_range Pool\_spa\_pumps Interior\_lighting Exterior\_lighting Lighting Plug Pool\_spa\_heater & The Datetime attribute has values in the format of YYYY-MM-DD HH:00:00.  & 3 &  The three sources are hourly datasets with 34, 151, and 32 attributes, respectively, mapped to 13 detailed end-uses. For example, the sum of electricity\_pool\_pump\_kwh' and 'electricity\_hot\_tub\_pump\_kwh in source-1 is mapped to Pool\_spa\_pumps in the target.\\

         \midrule

        11 & Monthly Hour-Level Load Transformation by Detailed Enduse  & Month Hour HVAC water\_heating Refrigerator Clothes\_washer Clothes\_dryer Dishwasher Cooking\_range Pool\_spa\_pumps Interior\_lighting Exterior\_lighting Lighting Plug Pool\_spa\_heater & Month is an integer number from 1 to 12. Hour is an integer number from 0 to 24. & 4 & Similar to Group 10, with one additional seasonal source dataset with an End Use Category attribute of which the values map to the 13 detailed end uses.\\

         \midrule

        12 & Seasonal Hour-Level Load Transformation by Detailed Enduse & Season Hour HVAC water\_heating Refrigerator Clothes\_washer Clothes\_dryer Dishwasher Cooking\_range Pool\_spa\_pumps Interior\_lighting Exterior\_lighting Lighting Plug Pool\_spa\_heater & Season has values such as Spring, Summer, Fall, and Winter. Hour" is an integer number from 0 to 24. & 4& Similar to Group 11. \\

         \midrule

        13 & Daily Hour-Level Load Transformation by High-Level Enduse  & Datetime\ HVAC Domestic\_water\_heating Major\_appliances Lighting Miscellaneous\_plug\_loads Total & Similar to Group 10, except that this target has fewer (higher-level) end-uses. & 3 & Similar to Group 10. \\

         \midrule

        14 & Monthly Hour-Level  Load Transformation by High-Level Enduse  & Month Hour HVAC Domestic\_water\_heating Major\_appliances Lighting Miscellaneous\_plug\_loads Total & Similar to Group 11, except that this target has fewer (higher-level) end-uses. & 4 &  Similar to Group 11.\\

         \midrule

        15 & Seasonal Hour-Level Load Transformation by High-Level Enduse  & Season Hour HVAC Domestic\_water\_heating Major\_appliances Lighting Miscellaneous\_plug\_loads Total & Similar to Group 12, except that this target has fewer (higher-level) end-uses. & 4 & Similar to Group 12. \\

        \bottomrule
    \end{tabular}
\end{table*}

\subsubsection{Other Benchmarks Used}
We also used two other benchmarks that go beyond the smart building data transformation for different purposes. One commercial benchmark consists of $16$ cases used by the Auto-Pipeline baseline. Since Auto-Pipeline code is not publicly available, we apply our proposed approach (without and with domain-specific knowledge) to the benchmark and compare the results. 

Another benchmark consists of four COVID-19 data transformation cases, which we used to validate further how well our methodology can generalize to other data transformation scenarios. It includes all four transformation cases observed in the Github commit history of a widely used real-world COVID-19 data repository maintained by John Hopkins University~\cite{covid19-jhu}. The attributes in the target data are \textit{(Province/State, Country/Region, Last Update, Confirmed, Deaths, Recovered)}, which represent the state-level COVID-19 statistics. The source schemas of the first two cases involve county-level data with different numbers of columns, and the latter two cases involve state-level data with different column names and different numbers of columns.

Overall, we have tested $125$ cases in three benchmarks, among which, $27$ cases involve attribute merging, $89$ cases involve attribute name changes, $32$ cases involve pivoting, $5$ cases involve attribute flattening, $50$ cases involve group-by and aggregation, $8$ cases involve join.

\subsection{Evaluation Metrics}

We report the following metrics in the experimental study:

\vspace{3pt}
\noindent
$\bullet$ Execution Accuracy: This metric is defined as the ratio of the number of correctly transformed cases to the total number of transformation cases. For each case, if the LLM can return the correct transformation query that passes the experimental validation tests as described in Sec.~\ref{sec:validation} within $5$ iterations, it is considered a correctly transformed case.

\vspace{3pt}
\noindent
$\bullet$  Column Similarity: We compute the similarity score for each column in the transformed dataset and its corresponding column in the ground truth target dataset (defined in Sec.~\ref{sec:validation}). As detailed in Sec.~\ref{sec:validation}, we compute a similarity score for each column. We further define the \textit{column similarity per case} as the average similarity scores of all target attributes in the case, the \textit{column similarity per group} as the average similarity scores of all cases in the group, and the \textit{overall column similarity} as the average similarity scores of all cases in all groups. The similarity score is set to zero for cases that fail to generate output data for similarity comparison.

\vspace{3pt}
\noindent
$\bullet$  Number of Iterations to Success: For each case, we record \textit{the number of iterations} used to achieve the correct response for the case. We record the \textit{average number of iterations to success} for all successful cases that achieved a column similarity score of $1.0$ within $5$ iterations in each group, and in all groups. The latter is termed as the \textit{overall number of iterations to success}.

\subsection{Experimental Setups}

We implemented the end-to-end workflow as illustrated in Fig.~\ref{fig:workflow} in a Python script that uses the ChatGPT-3.5-turbor-16K model. We did not present results on ChatGPT-4, because the corresponding OpenAI API has a limit of $4$K bytes for the total prompt-response size at this point, while this size is insufficient for a significant portion of real-world cases. For example,  tables in Group 10 to Group 15 have up to $152$ attributes, leading to a large prompt size.  We set the temperature to zero to avoid randomness for several reasons. First, a primary goal of this work is to evaluate the effectiveness of LLMs on data transformation tasks by using different types of initial prompts and the effectiveness of iterative prompt optimization. Random responses require additional methods (e.g., majority voting) for self-consistency, which will complicate the comparison. Second, setting the temperature to zero will achieve better quality results in most cases, according to a recent OpenAI article~\cite{openai-service}. All SQL codes are run on PostgreSQL database version 15.0 for validation. All descriptions for source and target attributes are obtained from a domain-specific database\footnote{The domain-specific database is maintained by co-author Liang Zhang. Some example information in the database is illustrated in Fig.~\ref{fig:domain-specific-db-example} and Fig.~\ref{fig:example1}}.


\subsection{Smart Building Data Transformation Results}

\subsubsection{Overall Results.} The zero-shot learning results for the smart building data transformation benchmark are illustrated in Tab.~\ref{tab:smart-building-zero-shot-results}. Using Prompt-3, our proposed $\tt SQLMorpher$ methodology achieved an execution accuracy of $\textbf{96}\%$, which is significantly higher than Prompt-1 and Prompt-2, which achieved an execution accuracy of $28\%$ and $36\%$, respectively. It demonstrated the importance of supplying domain-specific knowledge, particularly schema change hints, as part of the prompt to the LLM. The observation justifies the integration of the LLM with the domain-specific knowledge base and the schema mapping tools for data transformation pipelines.

\begin{table*}[t]
    \centering
    \scriptsize
    \caption{\label{tab:smart-building-zero-shot-results} Comparison of Execution Accuracy Using Different Prompt Templates with Zero-Shot Learning (Grp stands for Group) }
    \begin{tabular}{p{1.8cm}rrrrrrrrrrrrrrr}
        \toprule
        &Grp-1&Grp-2&Grp-3&Grp-4&Grp-5&Grp-6&Grp-7&Grp-8&Grp-9&Grp-10&Grp-11&Grp-12&Grp-13&Grp-14&Grp-15\\
        \midrule
        \#keywords avg.&15.6&18.7&19.6&19.6&16.4&16.3&24.2&26.7&23.7&13.7&20.3&28.0&5.0&25.3&31.5\\
        length avg.&1802&2826&1957&2023&1731&1713&1548&3239&1034&1712&2085&2365&1412&1732&1918\\
        \midrule
         \multicolumn{16}{c}{Prompt 1. Overall execution accuracy: 29/105 (28\%); Overall column similarity scores: 0.4; Overall iteration to success: 1.3}\\
        \midrule
        exec auc &6/10&2/10&4/10&6/10&3/10&2/10&6/10&0/10&0/3&0/3&0/4&0/4&0/3&0/4&0/4\\
        sim score avg. &0.7&0.5&0.6&0.6&0.3&0.3&0.8&0.0&0.0&0.6&0.3&0.0&0.3&0.0&0.0\\
        iter-to-succ avg.&1.2&1.0&1.0&1.3&1.0&1.0&2.0&-&-&-&-&-&-&-&-\\
            \midrule
         \multicolumn{16}{c}{Prompt 2. Overall execution accuracy: 38/105 (36\%); Overall column similarity score: 0.5; Avg iteration to success: 1.1}\\
        \midrule
        exec auc &6/10&6/10&7/10&6/10&3/10&2/10&8/10&0/10&0/3&0/3&0/4&0/4&0/3&0/4&0/4\\
        sim score avg. &0.7&0.6&0.7&0.6&0.4&0.3&0.9&0.0&0.4&0.8&0.7&0.2&0.3&0.1&0.0\\
        iter-to-succ avg.&1.0&1.4&1.2&1.0&1.0&1.0&1.2&-&-&-&-&-&-&-&-\\
                    \midrule
         \multicolumn{16}{c}{Prompt 3. Overall execution accuracy: 101/105 (96\%); Overall column similarity score: 0.96; Avg iteration to success: 1.2}\\
        \midrule
        exec auc &10/10&10/10&10/10&10/10&10/10&10/10&10/10&9/10&3/3&3/3&4/4&3/4&3/3&2/4&4/4\\
        sim score avg. &1.0&1.0&1.0&1.0&1.0&1.0&1.0&0.9&1.0&1.0&1.0&0.8&1.0&0.5&1.0\\
        iter-to-succ avg.&1.0&1.0&1.1&1.0&1.7&1.6&1.0&1.0&1.0&1.3&1.5&1.0&1.3&1.0&1.0\\

        \bottomrule
    \end{tabular}
\end{table*}

\subsubsection{Effectiveness of One-shot Learning.} For the four cases that failed with Prompt-3, we applied Prompt-4, Prompt-5, and Prompt-6 to check whether providing one demonstration example that involves a similar prompt and a correct response can improve the LLM response. 
The results are illustrated in Tab.~\ref{tab:one-shot}, which showed that using prompts that combine domain-specific knowledge and demonstration is capable of solving all four complicated cases that failed with Prompt-3.

\begin{table}[t]
    \centering
    \scriptsize
    \caption{\label{tab:one-shot} Average Column Similarity Score with One-shot Learning}
    \begin{tabular}{lp{1.5cm}p{1.5cm}p{1.5cm}}
        \toprule
      Cases Failed with Prompt-3&Prompt-1 +Demo & Prompt-2 +Demo & Prompt-3 +Demo\\
         \midrule
        Case   78  (in Group 8) &1.00& 1.00&1.00\\
        Case   92  (in Group 12) & 0.00&0.27&1.00\\ 
        Case 100 (in Group 14) &0.38&0.50&1.00\\ 
        Case 101 (in Group 14) &0.50&0.50&1.00\\ 

        \bottomrule
    \end{tabular}
\end{table}


\subsubsection{Effectiveness of the Iterative Optimization Process.} 
\label{sec:iterative-results}

Compared to Prompt-1 and Prompt-2, we have found that Prompt-3 can gain significantly more from iterative prompt optimization. 
When using Prompt-1, five cases in three groups, Group-1, Group-4, and Group-7, benefit from iterative prompt optimization, the average number of iterations being $1.2$, $1.3$, and $2.0$, respectively, as illustrated in Tab.~\ref{tab:smart-building-zero-shot-results}. Other groups either have all cases passed in one iteration or have all cases failed. When using Prompt-2, four cases in three groups, Group-2, Group-3, and Group-7, require more than one iteration to succeed, the average number of iterations being $1.4$, $1.2$, and $1.2$, respectively. When using Prompt-3, $10$ cases in six groups, require more than one iteration to succeed. It means that $9.5\%$ of total cases can benefit from iterative prompt optimization when using Prompt-3.

\subsection{Results on Benchmarks Beyond Smart Building.} 
\label{sec:non-smart-building}

First, we tested our approach on the COVID-19 benchmark. The results are illustrated in Tab.~\ref{tab:covid}, which showed that our proposed methodology resolves all four cases simply using the basic prompt (Prompt-1).

\begin{table}[t]
    \centering
    \scriptsize
    \caption{\label{tab:covid} Prompt Comparison for Covid-19 Benchmark \\ \#keywords avg.: \textbf{5}, length avg.: \textbf{277}}
    \begin{tabular}{lrrr}
        \toprule
       & Prompt-1 & Prompt-2 & Prompt-3\\
         \midrule
       exec auc & 4/4& 4/4 & 4/4\\
       sim score avg. &1.0&1.0&1.0\\
       iter-to-succ avg. &1.0&1.0&1.0\\
        \bottomrule
    \end{tabular}
\end{table}

Second, we also compared our proposed approach with the Auto-Pipeline approach, using its commercial benchmark~\cite{yang2021auto}. The results are illustrated in Tab.~\ref{tab:commercial}. It showed that our proposed methodology achieved perfect execution accuracy on all $16$ transformation problems in their benchmark only using the basic prompt, without additional domain-specific knowledge. The execution accuracy achieved by Auto-Pipeline on this benchmark is below $70\%$ ~\cite{yang2021auto}. The comparison implies that our approach has great potential to outperform state-of-the-art automatic data transformation tools.


\begin{table}[t]
    \centering
    \scriptsize
    \caption{\label{tab:commercial} Comparison to  Auto-Pipeline on Their Commercial Benchmark \\ \#keywords avg.: \textbf{8}, length avg.: \textbf{566}}
    \begin{tabular}{lrrrr}
        \toprule
       & Prompt-1 & Prompt-2 & Prompt-3 & Auto-Pipeline \cite{yang2021auto}\\
         \midrule
       exec auc & 13/16& \textbf{16/16} & \textbf{16/16} & 11/16~\cite{yang2021auto}\\
       sim score avg. &0.83&1.00&1.00 & -\\
       iter-to-succ avg. &1.00&1.00&1.00 &-\\
        \bottomrule
    \end{tabular}
\end{table}

\begin{figure} [t]
\centering
\includegraphics[width=0.47\textwidth]{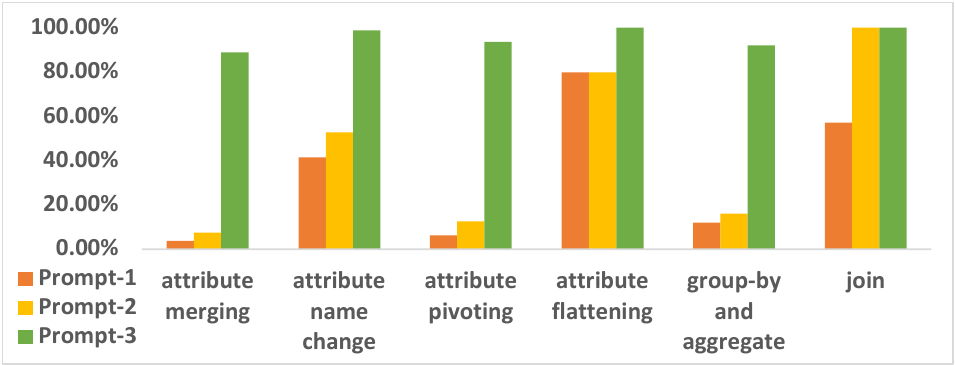}
\caption{\label{fig:groups-by-schema-changes} \small
The overall execution accuracy of cases in each schema change category. (We considered all $125$ cases in three benchmarks; each case may involve multiple types of schema changes.)
}
\vspace{-10pt}
\end{figure}

\subsection{Summary of Key Findings}
\vspace{3pt}
\noindent
$\bullet$  Large language models are promising to automatically resolve complicated smart building data transformation cases if domain-specific knowledge is available and easily retrievable. We achieved $96\%$ accuracy on our proposed benchmark, which consists of $105$ real-world smart building cases.

\vspace{3pt}
\noindent
$\bullet$  Our $\tt SQLMorpher$ methodology is promising in generalizing to other data transformation cases and outperforming state-of-the-art automatic data transformation tools that do not rely on LLMs. In particular, our methodology defines clean interfaces for integrating domain-specific knowledge into the data transformation process through the prompt generation process. This is a missing feature in state-of-the-art data transformation tools. The evaluation results on the commercial benchmark used by Auto-Pipeline showed that our approach, even without using any domain-specific knowledge, could achieve significantly better execution accuracy than Auto-Pipeline ($81\%$ vs. $69\%$). One observation is that while LLM generates SQL code, Auto-Pipeline attempts to learn a pipeline of data transformation operators. The latter has a more limited search space, which may affect the execution accuracy.

\vspace{3pt}
\noindent
$\bullet$  Compared to other domain-specific knowledge, a high-level schema change hint, such as column mapping relationships or instructions as simple as ``use aggregation", is critical to the success of our proposed methodology.

\vspace{3pt}
\noindent
$\bullet$ We further classify each of $125$ cases from all three benchmarks into one or more schema change types. We then count the execution accuracy for each type of schema change, as illustrated in Fig.~\ref{fig:groups-by-schema-changes}. We observe that while Prompt-3 with schema change hints can handle all schema change types well, Prompt-1 and Prompt-2 without schema change hints achieved relatively better accuracy ($40\%$ to $100\%$) for the attribute name change, attribute flattening, and join than other types of changes, such as attribute merging, attribute pivoting, and group-by/aggregation. This further verified the importance of incorporating high-level schema change hints such as ``use aggregation" and  ``use pivoting".

\vspace{3pt}
\noindent
$\bullet$ Zero-shot learning is effective in resolving most data transformation problems investigated in this work. Few-shot learning can resolve the difficult cases that fail with zero-shot.

\vspace{3pt}
\noindent
$\bullet$  The iterative optimization framework that simply enhances the prompt using ChatGPT reported errors or SQL execution errors for each iteration can benefit $9.5\%$ of cases when using Prompt-3 and $5\%$ of cases when using Prompt-1 and 2.

\vspace{3pt}
\noindent
$\bullet$  The examples in our proposed building energy data transformation benchmark are significantly more complicated than existing benchmarks in terms of the number of distinct keywords and the length of the transformation query. They are used in the real world but are missing in existing data transformation benchmarks~\cite{li2023auto, yang2021auto}.

\section{Conclusion and Future Works}
In this work, we pioneered the experimental and feasibility study about applying LLMs to data transformation problems. We proposed a novel $\tt SQLMorpher$  approach using LLM to generate SQL modification queries for data transformation. $\tt SQLMorpher$ is designed to incorporate domain knowledge flexibly and optimize prompts iteratively. We provided a unique benchmark for building energy data transformation, including $105$ real-world cases collected from $21$ energy companies in the United States.  The results are promising, achieving up to $\textbf{96}\%$ accuracy on the benchmark. In addition, we have found our system can generalize to scenarios beyond building energy data. The commercial benchmark results demonstrate that our approach is able to outperform existing automatic data transformation techniques significantly. In summary, $\tt SQLMorpher$ is promising to enable the automatic integration of diverse data sources for building energy management and may benefit other domains. In the future, we will design quality control for $\tt SQLMorpher$ to further reduce human validation involvement in the production environment.
\bibliographystyle{ieeetr}
\bibliography{refs}

\end{document}